\documentclass[runningheads]{llncs}
\usepackage{graphicx}
\usepackage{todonotes}

\begin{document}

\title{Immersive Technologies in Training and Healthcare: From Space Missions to Psychophysiological Research}

\titlerunning{Immersive Technologies in Training and Healthcare}

\author{
Barbara Karpowicz\inst{1,6}\orcidID{0000-0002-7478-7374} \and
Maciej Grzeszczuk\inst{1,6}\orcidID{0000-0002-9840-3398} \and
Adam Kuzdraliński\inst{1}\orcidID{0000-0003-2383-1950} \and
Monika Kornacka\inst{2}\orcidID{0000-0003-2737-9236} \and
Aliaksandr Marozau\inst{1,5}\orcidID{0000-0003-3800-2899}\and
Wiktor Stawski \inst{1,6}\orcidID{0000-0001-8950-195X} \and
Pavlo Zinevych \inst{1,6} \and
Grzegorz Marcin Wójcik \inst{3}\orcidID{0000-0003-2820-397X}\and 
Tomasz Kowalewski \inst{1,6}\orcidID{0009-0002-0542-9546} \and 
Grzegorz Pochwatko\inst{7}\orcidID{0000-0001-8548-6916} \and
Wiesław Kopeć \inst{1,4,6}\orcidID{0000-0001-9132-4171}
}

\authorrunning{Karpowicz et al.}

\institute{XR Center, Polish-Japanese Academy of Information Technology \url{https://xrc.pja.edu.pl}
\and
EC Lab, SWPS University of Social Sciences and Humanities 
\and
University of Maria Curie-Skłodowska, Lublin, Poland
\and
Kobo Association
\and
Proven Solution, Dubai, United Arab Emirates
\and 
XR Space \url{https://xrspace.xrlab.pl/}
\and
Institiute of Psychology of the Polish Academy of Sciences
}

\maketitle             

\begin{abstract}

Virtual, Augmented, and eXtended Reality (VR/AR/XR) technologies are increasingly recognized for their applications in training, diagnostics, and psychological research, particularly in high-risk and highly regulated environments. In this panel we discuss how immersive systems enhance human performance across multiple domains, including clinical psychology, space exploration, and medical education.

In psychological research and training, XR can offer a controlled yet ecologically valid setting for measuring cognitive and affective processes. In space exploration, we discuss the development of VR-based astronaut training and diagnostic systems, allowing astronauts to perform real-time health assessments. In medical education and rehabilitation, we cover procedural training and patient engagement. From virtual surgical simulations to gamified rehabilitation exercises, immersive environments enhance both learning outcomes and treatment adherence.


\end{abstract}
\section{Rationale}
\begin{figure}
    \centering
    \includegraphics[width=0.8\linewidth]{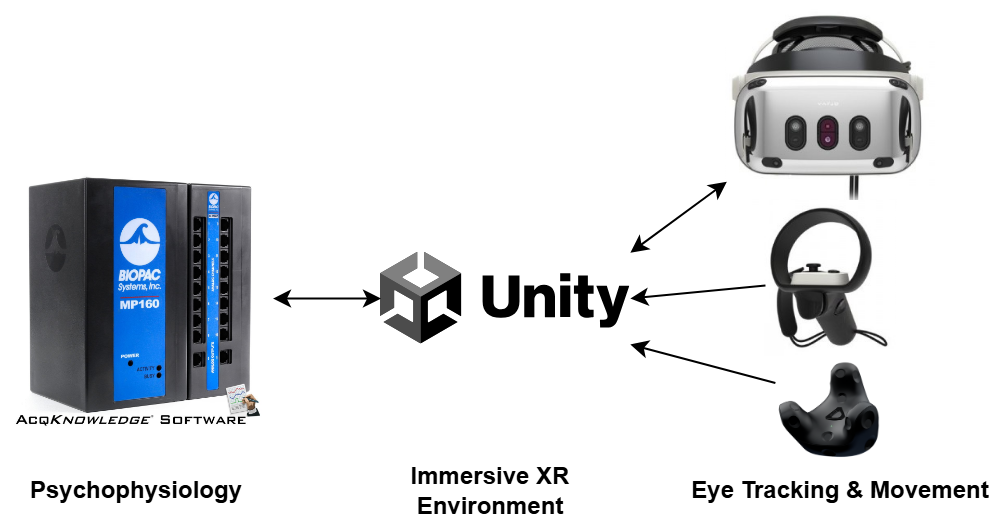}
    \caption{XR Framework Diagram (source: own elaboration)}
    \label{fig:xrFramework}
\end{figure}

Based on our previous research, including activities and projects related to Space Exploration (ALPHA-XR \cite{kopec2023human} and ESA Topical Team Space Analogs and Human Performance \cite{gabriel2024space}) we have successfully designed and implemented immersive systems that provide real-time acquisition of multimodal data across various XR-continuum immersive applications. The results have been used in numerous psychology research studies, giving valuable insights into user attention, cognitive load, and engagement in immersive environments.~\cite{karpo2022avatars,kopec2023human,pochwatko2023invisible,pochwatko2023wellbeing,schudy2023} As for today, our XR Framework consists of three parts: psychophysiology, immersive XR environment, and eye tracking, as depicted in Figure \ref{fig:xrFramework}.

A core focus of the first discussion point was on the implementation of XR Systems to \textbf{enhance human performance} by maintaining a balance between task difficulty and user engagement, often referred to as achieving 'flow', ensuring that the experience is neither too easy nor overly stressful.~\cite{csikszentmihalyi1992flow}

One of the challenges lies in implementing Dynamic Difficulty Adjustment (DDA) that works in real-time. Its implementation relies on a seamless integration of multimodal data that is collected, pre-processed and processed also in real-time. This data includes eye tracking that can be used, for example, to detect gaze patterns, area of interest (AoI), or changes in a pupil size. These objective metrics allow us to infer user states such as attention, cognitive load, and engagement without the need for subjective feedback, which could disrupt the task and add unnecessary stress. Another challenge is the performance of immersive systems. The challenge relies on the inherent architecture of modern Real-Time Computer Graphics (RTCG) systems, which in our case are responsible not only for rendering high-fidelity environments in real-time but also for the simultaneous multimodal data workflow and DDA.~\cite{karpo2024eyetracking} 

\section{Focus}

During the panel, we will discuss issues related to the design, creation, and implementation of immersive Virtual and eXtended Reality environments. A group of experts with a wide range of specializations, from artists through programmers, specialists in the field of human psychophysiology to doctors, will share their experiences with the application of modern technologies in scientific research, training, and medicine.

\subsection{Discussion Points}

\paragraph{Aspect 1: Technical challenges in implementing immersive environments for training and screening people.}

\paragraph{Aspect 2: eXtended Reality applications in the diagnosis and treatment of mental health disorders.}

\paragraph{Aspect 3: The role of virtual training environments in medicine, on the ground and in space applications.}

\section{Organizers}

\subsection{Panelists}

\paragraph{Barabara Karpowicz}
Computer scientist and researcher at the Faculty of Computer Science of the Polish-Japanese Academy of Information Technology (PJAIT). Vice-head of the XR Department and Chief Developer at the XR Research Center PJAIT (XRC PJAIT).
She is specializing in eXtended Reality (XR) technologies, human-computer interaction, and psychophysiology. Her work focuses on developing adaptive and user-centered Immersive XR Systems that integrate multimodal data such as eye tracking, body movement, and psychophysiological signals to enhance training, screening, and teleoperation.

\paragraph{Monika Kornacka}

Psychologist and researcher. Her research interests focus on emotion regulation, transdiagnostic processes (especially repetitive negative thoughts and divagation of thought), and cognitive-behavioral therapy. She is also interested in executive function training and the application of new technologies in clinical psychology and in research on psychopathology. She is a head of the Emotion Cognition Lab (ECL), where she uses the latest methods and technical solutions, such as mobile applications, eye tracking, virtual and extended reality, to measure psychological processes, and test how new technologies may be applied to psychological interventions.

\paragraph{Adam Kuzdraliński}

Biotechnologist and bioinformatician, combining interdisciplinary many fields of biological and technical sciences. Co-author of several patents, scientific publications (including in Nature Communications), popular science, R\&D project manager, entrepreneur. He co-created and managed biotechnology startups, and created technologies for their needs. Currently involved in the so-called AI red teaming in terms of biological threats generated by artificial intelligence systems, cooperating in this area with selected American entities.

\paragraph{Alexander Morozov}
Alexander Morozov is a physician with experience as a general practitioner and radiologist. He currently serves as Medical Officer and Subject Matter Expert, combining medical knowledge with modern technologies such as VR, AR, artificial intelligence, and robotics. He has led projects to develop products for medical education, support for people with autism spectrum disorders and other forms of neurodivergence, and pain management. These solutions are being used in various countries. Alexander Morozov is the author of 11 patents for inventions and utility models, as well as numerous scientific articles on various medical issues, including the combination of medicine with advanced technologies. He is actively involved in scientific research and projects linking future technologies to medical practice.

\paragraph{Wiktor Stawski}
Head of the XR\_C Students' Club and a researcher at the XR\_Center's laboratory. Passionate about 3D printing, Virtual Reality (VR), and eXtended Reality (XR). With a background in Interior Architecture, he has a keen interest in architecture and exhibitions. He presented a proposal for the reconstruction of an ancient Roman theater in Volterra using real-time engines at the MIDI 2024 conference. Additionally, he participated in an analog space mission at a simulation facility in collaboration with an ESA engineer. His work bridges the gap between traditional architecture and modern digital tools, showcasing innovative approaches to architectural implementation.

\paragraph{Wiesław Kopeć}
Computer scientist, research and innovation team leader, associate professor at Computer Science Faculty of Polish-Japanese Academy of Information Technology (PJAIT). Head of XR Center PJAIT and XR Department. He is also a seasoned project manager with a long-term collaboration track with many universities and academic centers, including University of Warsaw, SWPS University, National Information Processing Institute, and institutes of Polish Academy of Sciences. He co-founded the transdisciplinary HASE research group (Human Aspects in Science and Engineering) and distributed LivingLab Kobo.

\subsection{Researchers network}
We would like to thank the many people and institutions, gathered by the Living Lab Kobo and the HASE Research Group, to allow for this collaboration and research. Firstly, we thank all the members of HASE research group (Human Aspects in Science and Engineering) and Living Lab Kobo for their support. In particular, the members of the XR Research Center Polish-Japanese Academy of Information Technology (XRC PJAIT), including Mateusz Szewczyk, head of XRC Art\&Design Lab for visual identification system, as well as its XR\_C Students' Club, supporting ALPHA-XR efforts, the Emotion-Cognition Lab at SWPS University (EC Lab), VR and Psychophysiology Lab of the Institute of Psychology Polish Academy of Sciences, and Laboratory of Interactive Technologies (LIT) National Information Processing Institute (NIPI).

\section{Discussion and Conclusions}

To effectively design a training flow for complex, specialized tasks such as training, screening, teleoperation or piloting, it is crucial to consider the specifics of human performance. The task should not be too easy, but also not overly stressful. In order to limit cognitive load, we dynamically adjust task difficulty in real-time.
Gathering feedback solely in a declarative and subjective form — through direct responses from human subjects — is prone to error. Therefore, it is essential to supplement these signals with real-time biomonitoring data. Psychophysiological information, such as pupillometry or skin conductance, can provide insights into overload, even when the subject is unwilling to admit it or is unaware of their cognitive strain.
To maintain high realism in immersive simulation environments, it is crucial to sustain a high frame rate for generated visuals. Tasks that simultaneously tax the platform include not only scene construction and rendering of its objects but also the collection, processing, pre-processing, and evaluation of psychophysiological signals, all of which must occur in real-time.
We use eXtended Reality (XR), allowing the simulation participant to perceive their surrounding reality mixed with virtual elements. Complex immersive systems are particularly effective when real-world training would involve significant risks, either to life and health or to expensive equipment. Modern VR headsets enable high-resolution rendering, supporting activities such as reading and precise specialized tasks.

Virtual Reality (VR) and eXtended Reality (XR) technologies have traditionally been applied in clinical settings, particularly for treating patients with mental health disorders. However, beyond their therapeutic applications, these technologies offer valuable opportunities for assessing psychological processes in a controlled, yet ecologically valid manner.
A key challenge in psychological research is the reliance on self-report measures or laboratory experiments, both of which are often disconnected from the complexities of real-life experiences. Self-reports are inherently subjective and susceptible to biases, while laboratory-based experiments, despite their rigor, lack ecological validity. An alternative approach has been the use of mobile applications to track psychological processes in real-world contexts. However, this method introduces variability, since each participant's environment and situational factors differ significantly, making standardization difficult.
VR/XR offers a promising solution to these limitations by enabling researchers to measure psychological processes in immersive environments that closely resemble real-life settings while maintaining full experimental control. This allows for both ecological validity and standardization, ensuring that all participants engage in identical tasks under comparable conditions. Such an approach is particularly relevant for assessing emotion regulation, symptoms of psychopathology, and other clinically significant cognitive and affective processes in a way that is both reliable and applicable to clinical psychology.
Future research should explore how VR/XR-based assessments compare to traditional methods in terms of predictive validity and diagnostic accuracy. Additionally, integrating real-time biometric and behavioral monitoring into these immersive assessments could further enhance the precision of psychological measurements, paving the way for more personalized and effective mental health interventions.

In simulated, ground-based facilities, such as space-analogs, where participants operate under highly structured and regulated environments, individuals often face repetitive routines and isolation, which can negatively impact their well-being. Engaging in VR-based experiments introduced a meaningful psychological shift by providing a break from their daily tasks and offering varied, immersive experiences.

A key objective for the coming months is the development of a diagnostic system for space missions, enabling rapid and reliable health assessments in microgravity conditions. The foundation of this project is based on an expired patent related to polymerase chain reaction (PCR) technology, which presents an opportunity to develop a ready-to-use diagnostic system for trained astronauts. The process requires DNA extraction, amplification, and analysis, which necessitates both skilled operators and precisely regulated parameters. Given the constraints of space missions, including limited medical expertise onboard and the need for quick decision-making, training astronauts effectively is critical. Virtual Reality (VR) offers a promising solution for this challenge by providing immersive and standardized training modules that can prepare astronauts to perform DNA analysis and utilize diagnostic tools efficiently. While pre-prepared test kits can be used, proficiency in handling and interpreting results remains essential. VR/XR training ensures that crew members acquire the necessary skills before the mission, reducing the risk of errors in real-time applications. The ability to conduct rapid molecular diagnostics in space is crucial for identifying potential infections, determining pathology, and making timely medical decisions.

Virtual, Augmented, and eXtended Reality (VR/AR/XR) technologies offer numerous applications in the medical field, ranging from education and training to rehabilitation and patient engagement. Traditional medical education often presents challenges in spatial understanding, as students primarily rely on flat, two-dimensional images of organs and tissues. These representations can be limited, leading to difficulties in conceptualizing complex anatomical structures and physiological processes. In contrast, VR-based medical education allows for interactive three-dimensional visualization of organs, vessels, and systems, significantly enhancing spatial comprehension. Beyond anatomical understanding, VR also addresses social and psychological barriers in medical training. For example, auscultation training, which involves listening to heart sounds, can be challenging when students act as patients due to discomfort and distraction. In real clinical scenarios, both students and patients may experience hesitation or embarrassment, which can hinder effective learning. VR removes these social barriers, enabling learners to focus entirely on core clinical skills, leading to better knowledge retention and skill acquisition. For practitioners, especially surgeons, VR training provides substantial benefits. Surgical procedures require precise hand coordination, spatial awareness, and the ability to anticipate procedural steps. VR-trained surgeons have demonstrated improved procedural planning and execution due to enhanced visualization skills developed in immersive environments. Additionally, VR simulations facilitate team-based communication training, improving coordination among surgical staff, which is critical for patient safety. Unlike physical simulators, VR-based systems require less space, are cost-effective, and can be scaled for widespread adoption.

In rehabilitation, VR-based gamification enhances patient engagement and motivation. Traditional physical therapy exercises, such as repetitive limb movements, can be monotonous and demotivating for patients. By integrating gamified experiences — such as using a virtual sword to fight a dragon while performing prescribed movements — patients can maintain high engagement levels, improving adherence and therapeutic outcomes.

\section*{Acknowledgments}
Virtual and well-being environments used in studies have received funding from the EEA Financial Mechanism 2014-2021 grant no. 2019\slash 35\slash HS6\slash 03166.

\bibliographystyle{splncs04}
\bibliography{bibliography}

\begin{thebibliography}{1}
\providecommand{\url}[1]{\texttt{#1}}
\providecommand{\urlprefix}{URL }
\providecommand{\doi}[1]{https://doi.org/#1}

\bibitem{csikszentmihalyi1992flow}
Beck, L.A.: Csikszentmihalyi, mihaly.(1990). flow: the psychology of optimal experience (1992)

\bibitem{karpo2024eyetracking}
Karpowicz, B., Kowalewski, T., Zinevych, P., Kuzdrali{\'{n}}ski, A., Wójcik, G.M., Kope{\'c}, W.: Towards effective human performance in xr space framework based on real-time eye tracking biofeedback. In: "Digital Interaction and Machine Intelligence". Springer (2024)

\bibitem{karpo2022avatars}
Karpowicz, B., Mas{\l}yk, R., Skorupska, K., Jab{\l}o{\'{n}}ski, D., Kalinowski, K., Kobyli{\'{n}}ski, P., Pochwatko, G., Kornacka, M., Kope{\'{c}}, W.: Intergenerational interaction with avatars in vr: An exploratory study towards an xr research framework. In: Digital Interaction and Machine Intelligence. pp. 229--238. Springer, Cham (2022)

\bibitem{kopec2023human}
Kope{\'c}, W., Pochwatko, G., Kornacka, M., Stawski, W., Grzeszczuk, M., Skorupska, K., Karpowicz, B., Mas{\l}yk, R., Zinevych, P., Knapi{\'n}ski, S., et~al.: Human factors in space exploration: Opportunities for international and interdisciplinary collaboration. In: Machine Intelligence and Digital Interaction Conference. pp. 339--350. Springer (2023)

\bibitem{pochwatko2023invisible}
Pochwatko, G., Kopec, W., Jedrzejewski, Z., Jaskulska, A., Skorupska, K.H., Karpowicz, B., Masłyk, R., Barnes, S., Grzeszczuk, M., Lazarek, J., Swidrak, J.: The invisible – experienced: Developing and verifying a vr application for understanding air pollution perception and attitudes. In: 2023 IEEE International Symposium on Mixed and Augmented Reality Adjunct (ISMAR-Adjunct). pp. 531--536 (2023). \doi{10.1109/ISMAR-Adjunct60411.2023.00114}

\bibitem{pochwatko2023wellbeing}
Pochwatko, G., Kopec, W., Swidrak, J., Jaskulska, A., Skorupska, K.H., Karpowicz, B., Masłyk, R., Grzeszczuk, M., Barnes, S., Borkiewicz, P., et~al.: Well-being in isolation: Exploring artistic immersive virtual environments in a simulated lunar habitat to alleviate asthenia symptoms. In: Proceedings of the IEEE International Symposium on Mixed and Augmented Reality ISMAR. pp. 185--194. IEEE (2023). \doi{10.1109/ISMAR59233.2023.00033}

\bibitem{schudy2023}
Schudy, A., Pochwatko, G., Kopeć, W., Karpowicz, B., Skorupska, K., Grzeszczuk, M., Okruszek, {\L}.: A revised and extended paradigm for social and non-social stress elicitation in psychological research - a feasibility study in virtual reality. In: 2023 IEEE International Symposium on Mixed and Augmented Reality Adjunct (ISMAR-Adjunct) (2023)

\bibitem{gabriel2024space}
De~la Torre, G.G., Groemer, G., Diaz-Artiles, A., Pattyn, N., Van~Cutsem, J., Musilova, M., Kopec, W., Schneider, S., Abeln, V., Larose, T., et~al.: Space analogs and behavioral health performance research review and recommendations checklist from esa topical team. npj Microgravity  \textbf{10}(1), ~98 (2024)

\end{thebibliography}

\end{document}